\documentclass[12pt]{article}
\usepackage{color}
\usepackage{epsfig}
\usepackage{graphicx}
\usepackage{amssymb,amsmath}
\usepackage{afterpage}
\begin{document}
\pagestyle{plain}
\bibliographystyle{plain}

\vfill
\title{Neutrino mass hierarchy determination at reactor antineutrino experiments }
\vfill
\author{{ \small Guang Yang \footnote{Guang Yang: gyang9@hawk.iit.edu }} ~\\
\small on behalf of the JUNO collaboration \footnote{ANL is as an observer in JUNO currently.} ~\\
\small  Argonne National Laboratory, Lemont, Illinois 60439, USA ~\\
\small{\centering}and ~\\ 
\small Illinois Institute of Technology, Chicago, Illinois 60616, USA}
\date{August, 2015}
\maketitle

\begin{abstract}
After the neutrino mixing angle $\theta_{13}$ has been precisely measured by the reactor antineutrino experiments, 
one of the most important open questions left in neutrino physics is the neutrino mass hierarchy. 
Jiangmen Underground Neutrino Observatory (JUNO)~\cite{JUNO} is designed to determine the neutrino mass hierarchy (MH) without 
exploring the matter effect. The JUNO site location is optimized to have the best sensitivity for the mass hierarchy 
determination. JUNO will employ a 20 kton liquid scintillator detector located in a laboratory 700 meters underground. 
The excellent energy resolution will give us an unprecedented opportunity 
to reach a 3-4 $\sigma$ significance. In this paper, the JUNO detector design and simulation work will be presented. 
Also, RENO-50, another medium distance reactor antineutrino experiment, will do a similar measurement. With the efforts of these 
experiments, it is very likely that the neutrino mass hierarchy will be determined in the next 10 years.
\end{abstract}

\newpage
\section{\large MH and its determination}\

\begin{figure}
\centering
\includegraphics[angle=0,width=9cm,height=7cm]{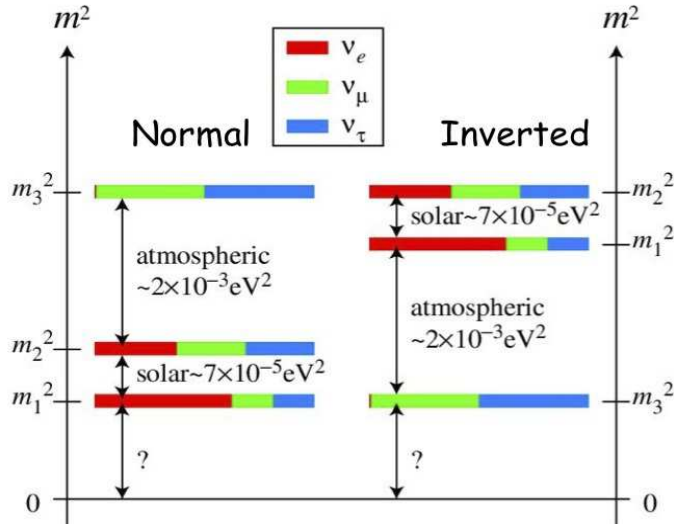}
\caption{{ The neutrino mass hierarchies.}}
\label{fig:fig_hierarchy}
\end{figure}

Recently, the reactor antineutrino experiments including Daya Bay, Double Chooz and RENO~\cite{DB,DC,RENO}
have successfully measured the value of the neutrino mixing angle $\theta_{13}$. This precise measurement 
stimulates the craving to solve the remaining neutrino puzzles in the neutrino community. One of them 
is the neutrino mass hierarchy determination. So far, we have already measured the sign of the neutrino mass
splitting $\Delta m^{2}_{21}$ but we do not know the sign of $\Delta m^{2}_{31}$ yet. So there are two 
neutrino mass hierarchy alternatives called normal hierarchy and inverted hierarchy. Fig.~\ref{fig:fig_hierarchy}
shows the two neutrino mass hierarchies. In current reactor antineutrino experiments, due to the tiny 
difference of $\Delta m^{2}_{31}$ between the two MH hypotheses and relatively large energy resolutions,
we are not able to determine the MH. However, in the next generation of reactor experiments, the
sensitivity to the MH will be improved significantly.
\

Currently, optimistic efforts to determine the neutrino mass hierarchy are basically accelerator-based 
long-baseline experiments like NOvA and T2K~\cite{NOVA,T2K}. They use muon neutrino or antineutrino
beams and try to detect the electron neutrino or antineutrino appearance which differ due to a matter effect. The different mass
hierarchies cause different appearance probabilities between neutrinos and antineutrinos in matter. 
However, a big disadvantage of these experiments is that the
mass hierarchy infomation is degenerate with the CP-violation phase.
In addition to the accelerator-based neutrino experiments, atmospheric neutrino experiments like
PINGU~\cite{PINGU} will attempt to determine the mass hierarchy using the matter effect as well.

\section{\large MH determination in medium-baseline reactor antineutrino experiments}\

\begin{figure}
\centering
\includegraphics[angle=0,width=10cm,height=6cm]{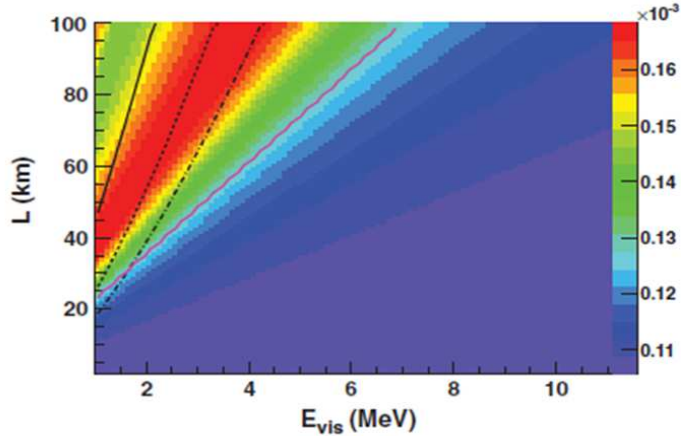}
\caption{{Oscillation differences between two mass hierarchies as functions
of the visible energy and baseline~\cite{Qian}.}}
\label{fig:fig_baseline}
\end{figure}

Medium-baseline reactor antineutrino experiments will have huge detectors. This kind
of experiment will simply use the mechanism that the neutrino oscillates in vacuum. 
Three such experiments have been proposed. They are Jiangmen underground neutrino observatory (JUNO) in China,
RENO-50 experiment in South Korea and WATCHMEN experiment in U.S. 
The electron antineutrino survival probability from reactors can be described as~\cite{Qian}:
\begin{equation}
\begin{split}
P(\overline{\nu}_{e}\rightarrow\overline{\nu}_{e})=1-2s^{2}_{13}c^{2}_{13}-4c^{2}_{13}s^{2}_{12}c^{2}_{12}\sin^{2}\Delta_{21} \\
+2s^{2}_{13}c^{2}_{13}\sqrt{1-4s^{2}_{12}c^{2}_{12}\sin^{2}\Delta_{21}}\cos(2\Delta_{32}\pm \phi),
\label{eq:eq_45}
\end{split}
\end{equation}
where $s$ stands for $\sin$, $c$ stands for $\cos$, $\Delta_{ij}\equiv|\Delta_{ij}|=1.27|\Delta m^{2}_{ij}|\frac{L [\rm{m}]}{E [\rm{MeV}]}$ and
\begin{equation}
\begin{split}
\sin\phi = \frac{c^{2}_{12}\sin(2\Delta_{21})}{\sqrt{1-4s^{2}_{12}c^{2}_{12}\sin^{2}\Delta_{21}}},
\label{eq:eq_46}
\end{split}
\end{equation}

\begin{equation}
\begin{split}
\cos\phi = \frac{c^{2}_{12}\cos(2\Delta_{21})+s^{2}_{12}}{\sqrt{1-4s^{2}_{12}c^{2}_{12}\sin^{2}\Delta_{21}}}.
\label{eq:eq_47}
\end{split}
\end{equation}
By using the above formulae, the baseline can be optimised in order to have the best sensitivity to the mass hierarchy.
Fig.~\ref{fig:fig_baseline} shows the differences between two mass hierarchies as
functions of the visible energy and baseline.   
Apparently, a detector with a distance over 40 km from the reactor cores will 
give us good sensitivity to the mass hierarchy.\

Another noticeable thing is that the precise determination of mass hierarchy relies on 
the knowledge of $\Delta_{32}$. Reference~\cite{Qian} has shown that
an incorrect $\Delta_{32}$ may bias the mass hierarchy determination to some extent.\

\section{\large JUNO experiment}\

\begin{figure}
\centering
\includegraphics[angle=0,width=15.7cm,height=4.5cm]{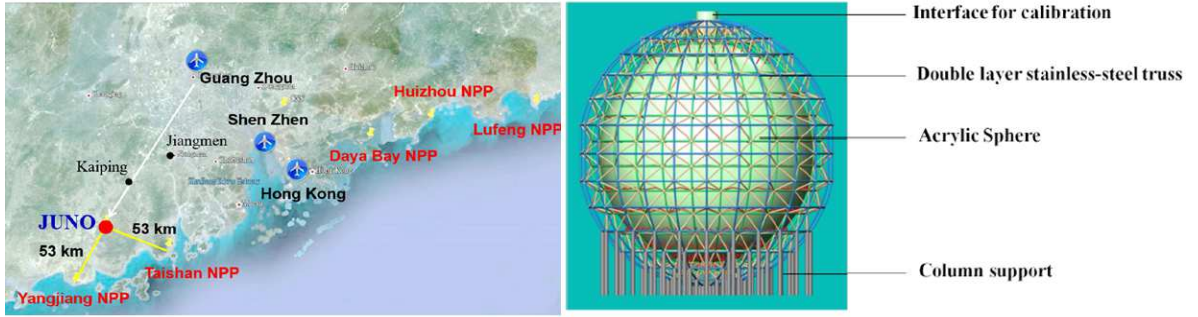}
\caption{{The left panel is the site of the JUNO experiment and the right panel
is the central detector design.}}
\label{fig:fig_JUNO1}
\end{figure}

The JUNO collaboration has been established for a few years and the civil construction is
on-going. JUNO will employ a 20 kton liquid scintillator detector and it is located
in Kaiping, Jiangmen, China. The detector has
a distance 53 km from the Yangjiang and Taishan nuclear power plants (NPP).
There are six 2nd generation pressurized water reactor (PWR) cores of 2.9 Gw$_{th}$ each (thermal power) in 
Yangjiang NPP and four 3rd generation PWR cores of 4.59 GW$_{th}$ each in Taishan NPP. However, the completion date of 
the reactors 3 and 4 in Taishan NPP is still unclear currently. 
In the absence of high mountains in the region where 
the detector has good sensitivity to the MH, the detector will be constructed
at a place 700 m deeply underground. There were multiple
options for the central detector design. The optimal one was that the central detector consists
of two concentric spherical tanks in a water pool. The outer one is made of stainless steel, in oreder to 
separate the shielding liquid from the water and the inner acrylic tank is filled with alklbenzene (LAB)
based liquid scintillator. Six kton of mineral oil will be filled into the outer stainless steel tank.
There will be about 15,000 20" photomultiplier tubes (PMTs) installed
on the inner wall of the steel tank. On the top of the water pool, the decommisioned detectors are used as muon trackers. They were
caloremeters for the OPERA experiment. Nonetheless, considering the difficulty of making two big tanks at once,
currently, the collaboration decides that the outer stainless steel tank is replaced by an 
open space truss or other steel supporting structures~\cite{JUNO2}. 
The left panel in fig.~\ref{fig:fig_JUNO1} shows the site of the JUNO detector and the reactors, and the right panel
shows the acrylic tank design of the central detector. More details can be found in~\cite{MIAO} and ~\cite{JUNO2}. 
\

\begin{figure}
\centering
\includegraphics[angle=0,width=14cm,height=5.5cm]{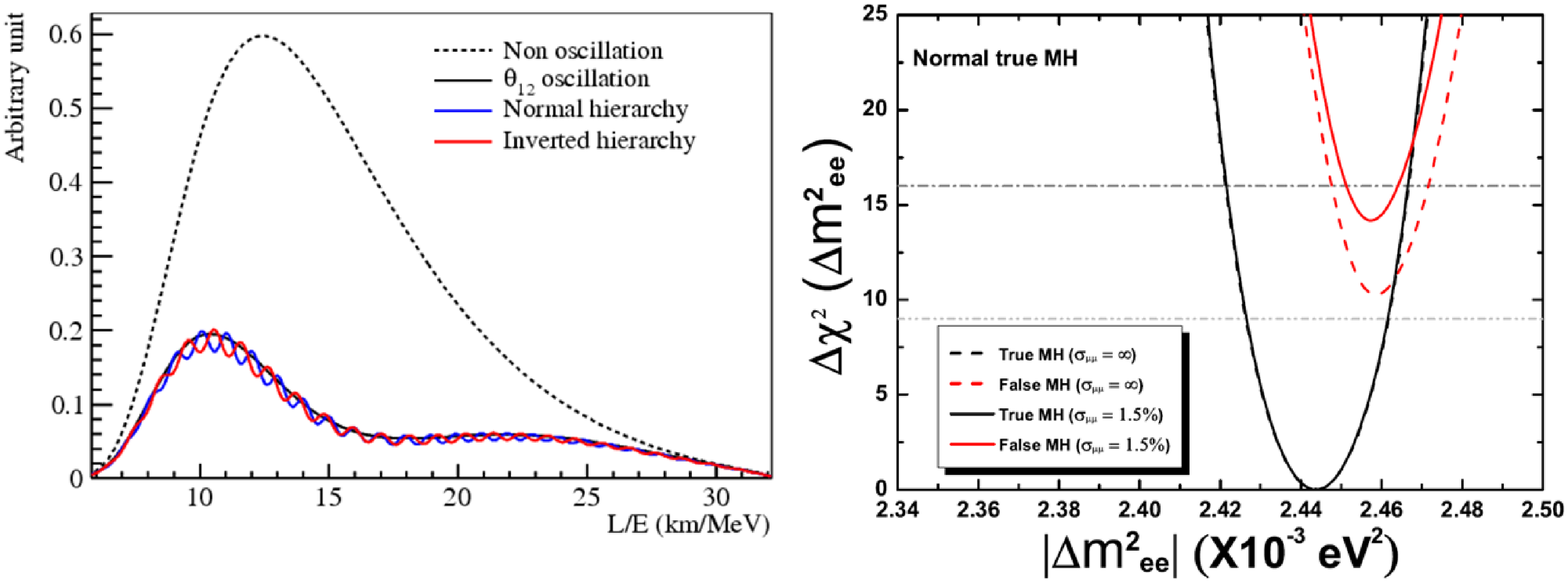}
\caption{{The left panel is spectra for two different mass hierarchies, non-oscillation and $\theta_{12}$ oscillation~\cite{JUNO} and the right panel
is the sensitivity for the mass hierarchy determination in JUNO~\cite{JUNO}.}}
\label{fig:fig_JUNO2}
\end{figure}

Similar to other reactor neutrino experiments, JUNO uses inverse beta decay to
detect neutrino signals. Since the difference of neutrino oscillation in vacuum 
for different mass hierarchies is very small, energy resolution is the
crucial factor for the success of JUNO. The goal is that the energy resolution 
can reach to 3$\%$/$\sqrt{E}$ at 1 MeV and the PMT coverage can reach about 80$\%$.
Furthermore, the PMT photocathode quantum efficiency will be greater than 35$\%$ and the attenuation length
of the liquid scintillator at 430 nm will be greater than 20 m~\cite{JUNO}. With these parameters,
the right panel of fig.~\ref{fig:fig_JUNO2} shows the sensitivity for the mass hierarchy
determination in JUNO while the left panel shows the energy spectra for different
mass hierarchies. There are two kinds of sensitivity studies performed. One is with
constraints of the neutrino mixing parameters from other experiments and the other one
is without any constraints on those parameters. They give confidence levels to 
reject the false mass hierarchy at more than 3 and 4 $\sigma$ significances. This sensitivity is obtained by taking into account
the reactor cores' distribution uncertainty, Daya Bay and Huizhou NPP contributions, spectrum shape uncertainty and the detector-related
uncertainties, including the energy nonlinear response of the detector. The background systematics, especially Li9 also have minor impact on the sensitivity~\cite{Jarah}. An assumption here is that
although Daya Bay~\cite{DB}, RENO~\cite{RENO} and Double Chooz~\cite{DC} found large discrepancies between data and prediction at 5 MeV region on the energy spectrum, model independent
prediction based on new precise measurements could improve the spectrum shape precision to 1$\%$~\cite{JUNO, Distortion}.
\

The civil preparation for JUNO has been done and the detector and underground lab are under construction.
The detector component production will be started in 2016 while the PMT production will be started at the 
same time. With hard effort, the detector will be assembled and installed in 2018 and 2019 potentially. If everything goes smoothly, the detector will
start data taking in 2020.

\section{\large Other efforts to the MH determination}\

In addition to the JUNO experiment, RENO-50 is proposed to employ a 18 kton liquid scintillator detector in South Korea, which has
a distance 47~km from the YG reactors. It is based on the success of the RENO experiment.
It will start the data taking from 2021 without any unexpected constraints. 
Potentially, it can give comparable MH sensitivities to the JUNO experiment~\cite{RENO50}. 
Besides, WATCHMEN experiment located in U.S. will use a water cherenkov detector
and it aims to determine the mass hierarchy~\cite{WATCHMAN}.

\section{\large Conclusion}\

Current accelerator and atmospheric neutrino experiments are making great efforts on 
the neutrino mass hierarchy determination by using the matter effect. Now reactor 
experiments like JUNO are proposed to determine the mass hierarchy precisely using 
neutrino oscillation in vacuum. They will have unprecedented size, PMT coverage and light
yield so that they have wonderful sensitivity to the mass hierarchy determination. In the next few decades,
the neturino mass hierarchy is very likely to be determined by the combination of these experiments.
Besides, experiments like JUNO can provide rich physics research opportunities other than the
mass hierarchy determination~\cite{JUNO, JUNO2}.

\end{document}